\PassOptionsToPackage{bookmarks=false}{hyperref}

\documentclass[sigconf]{acmart}
\renewcommand\footnotetextcopyrightpermission[1]{} 
\usepackage{booktabs} 
\usepackage{todonotes}
\usepackage{listings} 
\usepackage{enumitem}
\usepackage{graphicx}
\usepackage{subfigure}
\usepackage{caption}
\usepackage{tabularx,multirow,blindtext}
\usepackage{threeparttable}

\begin{document}
\title{PULP-HD: Accelerating Brain-Inspired High-Dimensional Computing on a Parallel Ultra-Low Power Platform}

   \author{Fabio Montagna$^\dag$, Abbas Rahimi$^{\ddag\star}$, Simone Benatti$^\dag$, Davide Rossi$^{\dag\star}$, and Luca Benini$^{\dag\star}$} 
  \affiliation{
  \institution{ $^\dag$University of Bologna, $^\ddag$UC Berkeley, $^\star$ETH Zurich}
  \city{ $\{$fabio.montagna, simone.benatti, davide.rossi$\}$@unibo.it, $\{$abbas, lbenini$\}$@iis.ee.ethz.ch}
  }

\begin{abstract}
Computing with high-dimensional (HD) vectors, also referred to as \textit{hypervectors}, is a brain-inspired alternative to computing with scalars.
Key properties of HD computing include a well-defined set of arithmetic operations on hypervectors, generality, scalability, robustness, fast learning, and ubiquitous parallel operations.
HD computing is about manipulating and comparing large patterns---binary hypervectors with 10,000 dimensions---making its efficient realization on minimalistic ultra-low-power platforms challenging.
This paper describes HD computing's acceleration and its optimization of memory accesses and operations on a silicon prototype of the PULPv3 4-core platform (1.5\,mm$^2$, 2\,mW), surpassing the state-of-the-art classification accuracy (on average 92.4\%) with simultaneous 3.7$\times$ end-to-end speed-up and 2$\times$ energy saving compared to its single-core execution. 
We further explore the scalability of our accelerator by increasing the number of inputs and classification window on a new generation of the PULP architecture featuring bit-manipulation instruction extensions and larger number of 8 cores.  
These together enable a near ideal speed-up of 18.4$\times$ compared to the single-core PULPv3.
 
\end{abstract}

\maketitle

\section{Introduction}
The brain's circuits are massive in terms of numbers of neurons and synapses, suggesting that large circuits are fundamental to the brain's computing~\cite{SDM,PlateTr,HD09}.
High-dimensional (HD) computing, aka \textit{hyperdimensional computing}~\cite{HD09}, is based on the understanding that brains compute with \emph{patterns of neural activity} that are not readily associated with numbers.
In fact, the brain's ability to calculate with numbers is feeble.
However, by virtue of the large size of brain's circuits, we can model neural activity patterns in points of a high-dimensional space, that is, with hypervectors.
When dimensionality is in the thousands (e.g., 10,000-D), the term ``hyperdimensional'' is used~\cite{HD09}.
Hypervectors are also holographic and (pseudo)random with independent and identically distributed (i.i.d.) components. 
Such hypervectors can be mathematically manipulated to not only classify but also to make associations, form hierarchies, and perform other types of cognitive computations~\cite{HD09}.
Key properties of HD computing include generality, scalability, a well-defined set of arithmetic operations on hypervectors, ubiquitous parallel operations, robustness, and graceful degradation making it possible to develop efficient nanoscalable learning machines~\cite{HD-TCAS17}.

HD computing is a complete computational paradigm that is easily applied to various learning problems, e.g., analogical processing~\cite{PlateTr}, language recognitions~\cite{IEDM_2016,RRAM-Computing_TCAS17}, and speech recognition~\cite{Rasanen2015con}. 
HD computing has also been used for multimodal data fusion and prediction, including categorization of body physical activities from several heterogeneous sensors~\cite{Rasanen14}, predicting behavior of mobile-device users (e.g., media player prediction)~\cite{Rasanen15}, and reactive robot learning~\cite{HD-ReactiveRobot}. 
More recently, HD computing has shown promise in biosignal processing and classification of raw electromyography (EMG)~\cite{EMG-HD} as well as electroencephalography (EEG)~\cite{BICT17} data with minimal information: e.g., in the absence of domain expert knowledge and using smaller training datasets, without affecting the robustness of classification.

At its very core, HD computing is about manipulating and comparing large patterns stored in memory.
Its mathematical operations allow a high degree of parallelism by needing to communicate with only a local component or its immediate neighbors.
Other operations such as distance computation can be performed in a distributed fashion.
However, there has been no silicon proof yet to assess the advantages of HD computing for a large extensive application. 
Hence, we target accelerating HD computing on a parallel ultra-low power (PULP) platform~\cite{PulpWebsite} which exploits near-threshold operation coupled with parallel execution over multiple cores. The architecture is described in detail in \cite{rossi2015pulp}.
This paper makes the following contributions:
\vspace{-1mm}
\begin{itemize}[wide=0.5pt]
\item We present an accelerator for all operations of HD computing and optimize their memory accesses on a PULP platform. 
We target a silicon prototype of the PULP platform featuring 4 cores operating at 0.5\,V, fabricated in 28\,nm FD-SOI technology aka PULPv3~\cite{PULPv3}.
To the best of our knowledge, this is the first realization of an accelerated HD computing on an embedded platform with tight resources (1.5\,mm$^2$, 2\,mW) fabricated in the standard silicon-based technology.
We efficiently represent the components of binary hypervectors to unsigned integer arrays and carefully optimize their layout in L1/L2 memory; this enables double buffering for efficient data transfer and naturally exploits data level parallelism with bitwise and distributed operations. 
These optimizations construct a \emph{universal} accelerator for all applications of HD computing that are described using the open multiprocessing (OpenMP) directives.
This paper shows an example of using this accelerator in the EMG-based hand gesture recognition for a highly energy-efficient and wearable form-factor system.
\item Our accelerator preserves the semantic of HD computing by avoiding any lossy optimization on binary hypervectors, and its classification accuracy (on average 92.4\%) matches the golden MATLAB model\footnote{MATLAB code, C code for ARM Cortex M4 and our accelerator are open access at https://github.com/fabio-montagna/PULP-HD}.
This classification accuracy already surpasses the state-of-the-art support vector machines (SVMs)~\cite{benatti2017prosthetic}.
We demonstrate that HD computing is computationally affordable on a mW platform, and highly amenable for perfect parallel execution: PULPv3 with 4 cores achieves 3.7$\times$ end-to-end speed-up and 2$\times$ energy saving compared to its single core execution.
\item We further investigate how acceleration of HD computing can benefit on a new generation of the PULP featuring RISC-V based processors (called Wolf) extended for energy-efficient digital signal processing~\cite{gautschi2017near} such as bit-manipulation instructions.
This instruction extension together with a larger number of 8 cores achieves 18.4$\times$ speed-up compared to the single-core PULPv3.
We also evaluate the scalability of our accelerator by increasing the number of input channels and larger temporal windows of classification. 
We observe that HD computing scales very well, and the savings linearly benefit from a large number of cores paving the way for the development of future HD-centric accelerators.

\end{itemize}  
 
\section{Background}
In this section, we first provide background in HD computing, and describe the main modules of an HD computing-based classifier.
Then, we describe in details the PULP platform that is used for acceleration of the HD classifier.
\subsection{High-dimensional (HD) Computing}
Computing with 10,000-bit words takes us into the realm of very high-dimensional spaces and vectors.
There exist a huge number of different, nearly orthogonal hypervectors with the dimensionality in the thousands~\cite{SDM,HD09}.
This lets us combine two such hypervectors into a new hypervector using well-defined vector space operations, while keeping the information of the two with high probability.
The binary hypervectors are initially taken from a 10,000-D space and have an equal number of randomly placed $1$s and $0$s, i.e., $\lbrace 0, 1\rbrace^D$~\cite{BSC96}.
The number of places at which two binary hypervectors differ is called the Hamming distance and it provides a measure of \emph{similarity} between hypervectors.

HD computing uses three operations: multiplication, addition, and permutation (MAP).
The addition of binary hypervectors $[A + B + \ldots]$ is defined as the componentwise majority with ties broken at random.
The multiplication is defined as the componentwise XOR ($\oplus$), and permutation ($\rho$) shuffles the components, e.g., 1-bit rotation. 
All these MAP operations produce a D-bit hypervector.
The usefulness of HD computing comes from the nature of the operations.
Specifically, the addition produces a hypervector that is \emph{similar} to the input hypervectors, whereas multiplication produces a \emph{dissimilar} hypervector.
Hence, the addition is well suited for representing sets, and the multiplication is useful for binding two hypervectors.
The permutation also generates a dissimilar pseudo-orthogonal hypervector that is good for storing a sequence.
The multiplication and permutation are invertible.
\subsubsection{Modules of HD Classifier}
\label{section:ComponentsHD}
In the following, we describe three main modules for classification using HD computing.
First, an item memory (IM) maps all symbols in the system to the HD space.
In a typical biosignal processing system, the names of channels (or electrodes) are the basic symbols for mapping.
The IM assigns a random hypervectors (with i.i.d. components) to every channel's name, i.e., $E_1 \perp E_2 ... \perp E_{i}$.
Besides the discrete symbols, the system has analog values (e.g., the signal levels of channels) for mapping.
To map these analog values, the notion of IM is further extended to a \emph{continuous} item memory (CIM)~\cite{EMG-HD}.
In the continuous vector space of CIM, orthogonal endpoint hypervectors are generated for the minimum and maximum signal levels.
For instance, when the channel $i$ produces a maximum signal level at time $t$ and the minimum signal level at $t+k$, the corresponding generated hypervectors by CIM are orthogonal, i.e., $V_i^t \perp V_i^{t+k}$.
The hypervectors for intermediate levels are then generated by linear interpolation between these endpoints and are prestored in the CIM~\cite{EMG-HD,BICT17}.
The IM and CIM stay fixed throughout the computation, and they serve as seeds from which further representations are made.

Second, the seed hypervectors are encoded by the MAP operations to represent the event of interest for classification.
For instance, a \emph{spatial} encoder can represent a set of all channel-value pairs at timestamp $t$ into a binary hypervector ($S^t$).
To this end, the multiplication is used to bind each channel to its signal level, and to form the set all these bound hypervectors are bundled by the addition, i.e., $S^t=[(E_1 \oplus V_1^t) + ... + (E_i \oplus V_i^t)]$.
The generated binary hypervector ($S^t$) only captures the spatial information for a given time-aligned samples of channels.
However, in many applications \emph{temporal} information is of concern as well.
A temporal encoder can capture the relevant temporal information by using the permutation and multiplication that together form an N-gram hypervector from a sequence of N hypervectors .
Hence, a sequence of N spatial hypervectors at consecutive timestamps are encoded into an N-gram hypervector: 
$S^t \oplus \rho^{1} S^{t+1} \oplus \rho^{2} S^{t+2} \oplus... \oplus \rho^{n-1} S^{t+n-1}$ where $\rho^{k}$ is a rotation over $k$ positions of the hypervector.
Some biosignal processing applications such as EEG-based brain-machine interfaces may require a large temporal window as large as N-gram of 29~\cite{BICT17}. 

Finally, for a given class, across all its trials, the corresponding N-gram hypervectors are added to produce a binary \emph{prototype} hypervector.
During training, the prototype hypervectors are stored in an associative memory (AM) as the \emph{learned} patterns.
During classification, in an identical way to prototypes, a \emph{query} hypervector is generated from unseen inputs.
The AM compares the query hypervectors to all learned prototype hypervectors, and returns the label of the one that has the minimum Hamming distance.
Since these three modules are commonly used across various applications of HD computing~\cite{HD-TCAS17,EMG-HD,BICT17}, we target their accelerations to achieve end-to-end benefits in learning and classification tasks.    
\subsection{Parallel Ultra-Low Power (PULP) Platform}
The PULPv3 SoC used is in work exploits a software programmable, 4 processors cluster architecture operating in near-threshold (0.7\,V--0.5\,V), fabricated in 28\,nm FD-SOI technology~\cite{PULPv3}. 
The processors used in the cluster are based on an optimized implementation of the open-source OpenRISC instruction set architecture (ISA) which share 48\,kB of multi-banked tightly coupled data memory (TCDM) acting as software-managed L1 scratchpad memory. 
The 64\,kB off-cluster L2 memory can be accessed by a tightly coupled direct memory access (DMA) optimized for low power through the 64-bit AXI4 interconnect, which guarantees high L1 to L2 communication bandwidth (i.e., up to 32 Gbit/s at 500\,MHz). 

The cluster and the rest of the SoC (which includes L2 memory and peripherals) reside in two clock and power domains controlled by frequency-locked loops (FLLs), and external voltage regulators~\cite{PULPBoard}.
Hence, voltage and frequency can be scaled according to the performance requirements of the applications. 
The SoC features a standard peripheral set which includes SPI, QSPI, UART, I2C, I2S to connect to external commercial devices such as analog to digital converters (ADC). 
PULPv3 relies on OpenMP v3.0, as the de facto standard parallel programming model, that operates on top of GCC 4.9 toolchain.
The OpenMP implementation is based on a highly optimized bare-metal library to exclude an operating system that would otherwise introduce huge software overheads not suitable for ultra-low-power parallel accelerators. 

\section{Accelerating HD Computing on PULP}
\label{Section:acceleration}
This section describes the acceleration of the HD Computing on the PULP platform.     
To reduce the computational requirements, we first pack a binary hypervector in an array of conventional data type.
We directly map 32 consecutive binary components of a hypervector to an unsigned integer variable with 32 bits. 
In this way, a binary hypervectors, with 10,000 randomly placed 1s and 0s, can be losslessly represented with 313 unsigned integers. 
This leads to a significant reduction of the memory accesses. 
Moreover, having hypervectors united in 32 bits unsigned integers paves the way to aggressive code optimizations for the MAP operations, using simple componentwise majority, XOR, and shift operations.  

\begin{figure}[t]
\includegraphics[width=0.49\textwidth]{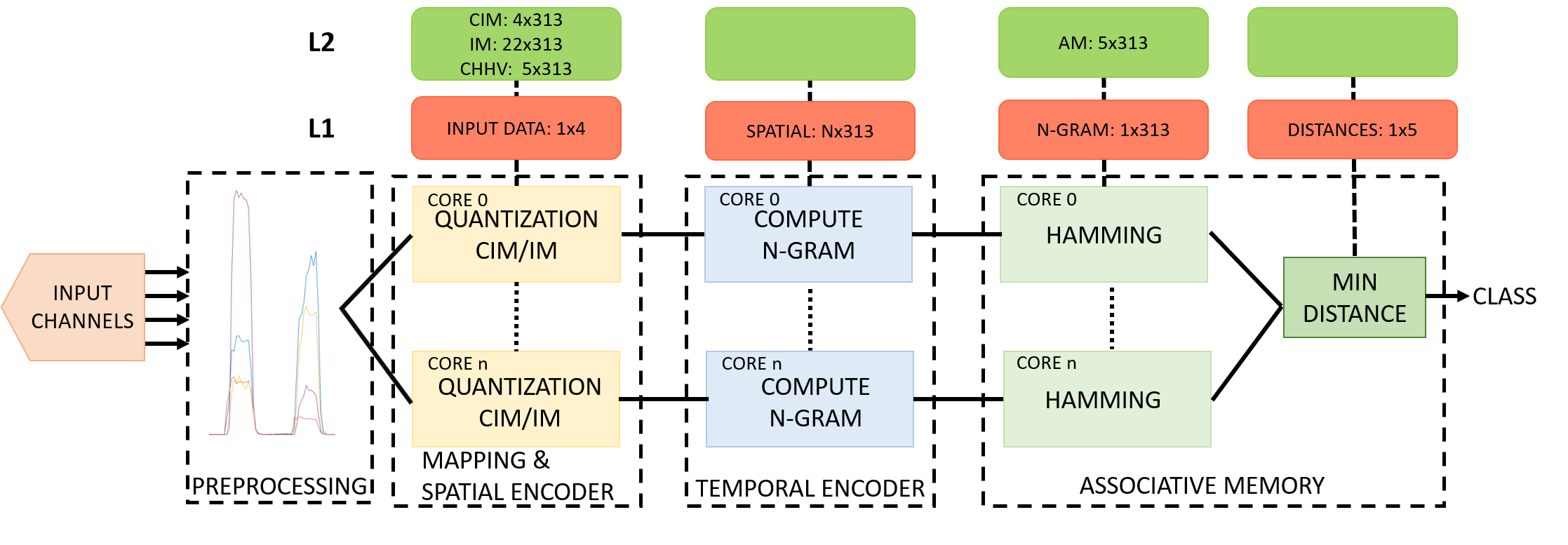}
\caption{Parallel processing chain of HD computing. The number of cores $n$ varies from 1 to 8 depending on the target architecture. Empty boxes indicate no memory usage.}
 
\label{fig:proc_chain}
\vspace{-5mm}
\end{figure}

Fig.~\ref{fig:proc_chain} illustrates the processing chain of HD computing that is composed by three main kernels: mapping to the HD space and spatial encoder, temporal encoder, and AM for classification.
Each kernel of the processing chain is parallelized separately using an optimized version of the OpenMP directives to efficiently distribute the workload over multiple cores.

The input EMG signals are acquired through a 16-bits ADC~\cite{EMG_SVM} and casted to  the 32-bit floating-point representation. The preprocessing block includes power line interference removal and envelope extraction.
This preprocessing block is not executed on the PULP platform, hence we exclude it from our parallel processing chain. %
The first kernel of our processing chain maps the EMG samples to the HD space, and performs spatial encoding among the channels.
To map the four samples (see Fig.~\ref{fig:proc_chain}) to the internal HD representation, we use the CIM with a fixed number of levels, e.g., 22 linear levels are suitable for the EMG task where the amplitude of signal typically ranges from 0 to 21\,mV.
A set of 22 hypervectors corresponding to each of these levels are generated offline and prestored in the CIM. 
The CIM utilizes a simple quantization step in which every sample is rounded to the closest integer level. 
Besides, the 4 EMG channels are also mapped to corresponding hypervectors on the IM with 4 orthogonal hypervectors.
In this kernel, parallelization is performed at data level. 

\begin{figure}[t]
\includegraphics[width=1\columnwidth]{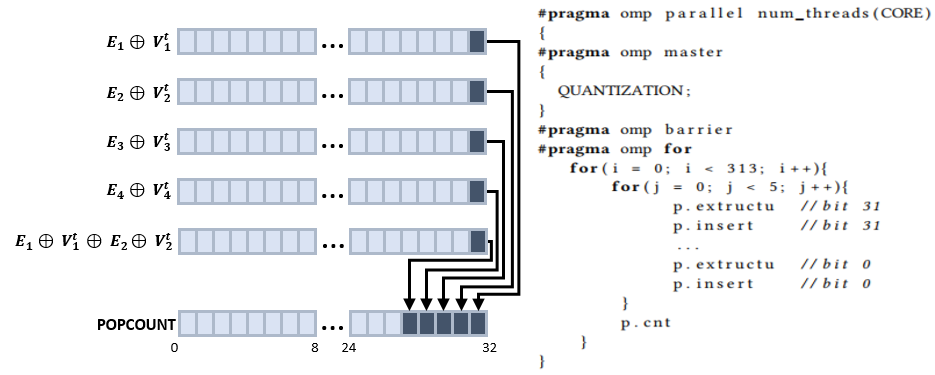}
\caption{Illustration of how built-ins (p.extractu, p.insert, p.cnt) are used in the spatial encoder (left). A code snippet to show how parallelism is achieved through OpenMP directives in the entire processing chain (right).}
\label{fig:popenc}
\vspace{-5mm}
\end{figure}

After mapping to the HD space, the workload is equally distributed among the cores, giving to each core a portion of the hypervectors on which the required encoding operations are performed. 
In this way, the cores execute first the componentwise XOR operation from the outputs of CIM and IM, and then the componentwise majority to create the spatial hypervector in parallel.
In Fig.~\ref{fig:popenc} (right), a code snippet of the spatial encoder kernel is presented to show how OpenMP directives are used for the parallelization of the entire processing chain.
In Section \ref{section:discussion}, we demonstrate that this kernel shows an high level of scalability on large number of cores, achieving a nearly ideal speed-up. 

The spatial hypervector (1x313), which goes as input in the temporal encoder, is stored directly in the L1 memory to avoid useless accesses to the high latency memory (L2) and requires 2\,kB of memory. 
Instead, the CIM (22x313 matrix, 27\,kB) and IM (4x313 matrix, 5\,kB) are stored in the L2 memory. 
By applying a double buffering policy via DMA, data are moved from high latency memory (L2) to L1 memory while the cores are processing the data already available in L1.
In this way, data transfers and processing phases can be superimposed improving the performance and the energy efficiency of the system.

As mentioned in the Section \ref{section:ComponentsHD}, in the case of N-grams greater than one, after the spatial encoder, the temporal encoder is executed.
In this kernel, a sequence of N spatial hypervectors are encoded and combined through a componentwise XOR operation after shifting them by one position as permutation. 
The output of this kernel is the N-gram hypervector that requires 2\,kB (stored the in L1 memory), and serves as the input of the AM kernel.

The last part of the processing chain is the AM kernel. 
The AM matrix is composed by the prototype hypervectors associated to each classes, derived from the learning session performed off-line.
Nevertheless, the AM matrix can be continuously updated for on-line learning. 
The Hamming distances are calculated between the query hypervector and all of hypervectors contained in the AM matrix. 
This kernel is parallelized at data level as well: the hypervectors are equally distributed among the cores to perform componentwise XOR between the components of the query and the components of the AM, and count the number of mismatches as distances.
The AM (5x313) matrix requires 7\,kB and is allocated in the L2 memory.
Here as well, the data are efficiently transferred from the L2 to L1 memory through the double buffering via the DMA.
The total memory requirements for the EMG application, considering 10,000-D hypervectors is around 50\,kB, perfectly matching the storage capabilities of the PULPv3 SoC.

\section{Experimental Setup and Results}
\label{section:system_description}

We evaluate our accelerator on an EMG dataset acquired from five subjects~\cite{EMG-HD}.
The EMG signals are sampled at 500\,Hz from four channels placed on the forearm of the subjects.
The dataset includes four common hand gestures: closed hand, open hand, 2-finger pinch, and point index. 
It also includes the rest position between subsequent gestures. 
The gesture are three second long and each one is repeated 10 times.
\subsection{Comparison with SVM on ARM Cortex M4}
In literature, the most used algorithms for EMG gesture recognition are support vector machine (SVMs), linear discriminant analysis (LDA) and k-nearest neighbor (KNN).  As shown in \cite{oskoei2008support}, through these techniques, SVM yields the highest accuracy.
Hence, we first show a comparison between classification accuracy of HD computing and the state-of-the-art SVM~\cite{benatti2017prosthetic}.
Then, we measure performance and power consumption of specialized serial versions of these two algorithms on a commercial embedded ARM Cortex M4, featuring the most commonly used ISA in the low-power embedded domain.

As the first step, we implement and validate these two algorithms on MATLAB to establish a golden model to follow. 
We use an identical setup for both algorithms as presented in~\cite{EMG-HD}; the model training is done per subject and off-line using 25\% of the dataset, while the entire dataset is used for testing.
The mean classification accuracy of gestures among five subjects is 89.6\% with SVM, and 92.4\% with the HD classifier. 
More importantly, the HD classifier exhibits a graceful degradation with lower dimensionality, or faulty components, allowing a trade-off between the application's accuracy and the available hardware resources in a platform~\cite{EMG-HD,HD-TCAS17}.  
We can exploit this graceful degradation capability by reducing the dimensionality of hypervectors that eases the execution on the commercial ARM Cortex M4.
%
%
%
%
%
To do so, we perform simulations by reducing the dimensionality from 10,000 to 100.  
The HD classifier closely maintains its accuracy when its dimensionality is reduced from 10,000 to 200, but beyond this point the accuracy is dropped significantly.
Therefore, for this experiment, we fix its dimension to 200-D showing a mean accuracy of 90.7\%, slightly higher than the SVM. 
This tuning allows compacting a hypervector to seven unsigned integers, and linearly reduces the number of operations of the HD classifier with no significant impact on its accuracy (i.e., iso-accuracy with SVM). 

On the other hand, the SVM does not support such a flexibility.
A trained model of SVM is composed by a number of support vectors (SVs). 
This parameter is not determined \emph{a priori}, and can vary due to several factors such as the scaling of the data, the kernel function, and the level of tolerable miss-classification. 
Obviously, all this variability requires time to find the best configuration that leads to the smallest number of SVs maintaining the highest accuracy. 
Hence, a different number of SVs and the dimension of input feature vectors (i.e., the number of channels) induce substantial differences in performance.
For this exploration, the dimension of the SVs is fixed to four as the number of input channels, while the number of SVs varies significantly across the model of five subjects, and finally is chosen to be 55 as the smallest among the subjects.
This is in sharp contrast to the HD classifier since there is no variability in its model size after choosing its parameters: the dimension of the hypervectors, the N-gram size, and the number of input channels.
%
\begin{table}[t]
  \caption{Comparison of HD computing (200-D) versus SVM at iso-accuracy on ARM Cortex M4. The results refer to a 10\,ms detection latency.}
  
  \label{tab:stmVSpulpV3}
  \scalebox{0.90}{
  \begin{tabular}{lcc}
    \toprule
  & \multicolumn{2}{c}{\textbf{ARM Cortex M4}} \\ 
    \midrule
    \textbf{Kernel} 		& Cycles(k) &Accuracy(\%)\\
    \texttt{HD COMPUTING}       &12.35 	&90.70    \\
    \texttt{SVM}  				&25.10 	&89.60   \\
    \bottomrule
  \end{tabular}
  }
\vspace{-4mm}
 
\end{table}  
Table~\ref{tab:stmVSpulpV3} summarizes the performance and accuracy results derived from the serial execution of the two algorithms on the ARM Cortex M4. 
The CIM, IM, and AM matrices of the HD classifier, and the SVs and coefficients matrices of the SVM, as the trained models, are loaded into the ARM Cortex M4 for testing.
For SVM, a fixed-point approach is used to avoid all the computation needed to be executed in the floating-point.
It is already demonstrated~\cite{montagna2017flexible} that this approach leads to best performance preserving the accuracy.
As shown, the HD classifier achieves $\approx$2$\times$ faster execution and lower power at iso-accuracy compared to the SVM on the ARM Cortex M4.
This is due to the fact that HD classifier mostly uses basic componentwise operations on the hypervectors.
In the following, we show how HD classifier can further benefit from our accelerator.
\subsection{HD computing on PULPv3 versus ARM Cortex M4}
\begin{table}[t]
  \caption{Detailed power (P) comparison of HD algorithm on the ARM Cortex M4 and PULPv3 based on number of cycles (CYC) and frequency (FREQ). The results refer to a 10\,ms detection latency.}  
  \label{tab:stmVSpulpV3_energy}
  \scalebox{0.64}{
  \begin{tabular}{l|c|c|c|c|c|c|c}
    \toprule
    & \multicolumn{1}{|c|}{\textbf{CYC}} & \multicolumn{1}{|c|}{\textbf{FREQ}} & \multicolumn{1}{|c|}{\textbf{FLL P}} & \multicolumn{1}{|c|}{\textbf{SOC P}} & \multicolumn{1}{|c|}{\textbf{CLUSTER P}} & \multicolumn{1}{|c|}{\textbf{TOT. P}} & \multicolumn{1}{|c}{\textbf{P BOOST}} \\
   & \multicolumn{1}{|c|}{\textbf{[k]}}   & \multicolumn{1}{|c|}{\textbf{[MHz]}} & \multicolumn{1}{|c|}{\textbf{[mW]}} & \multicolumn{1}{|c|}{\textbf{[mW]}} & \multicolumn{1}{|c|}{\textbf{[mW]}} & \multicolumn{1}{|c|}{\textbf{[mW]}} & \multicolumn{1}{|c}{\textbf{[$\times$]}} \\
    \midrule
    \hline
    \textbf{HD COMPUTING}   &	&   & 	&    &     &  \\
    \texttt{ARM CORTEX M4@1.85V}  &439     &43.90 &- &20.83 	&N.A. &20.83  &-     \\
    \hline
    \texttt{PULPv3 1 CORE@0.7V}    &533    &53.30 &1.45 &0.87  &1.90	&4.22  &4.9     \\

   \texttt{PULPv3 4 CORES@0.7V}     &143 	&14.30 &1.45 &0.23 &0.88 &2.56 &8.1 \\
   \texttt{PULPv3 4 CORES@0.5V}     &143	&14.30 &1.45 &0.23 &0.42 &2.10  &9.9 \\
    \bottomrule
  \end{tabular}
 }
 \vspace{-4mm}
\end{table}

Table~\ref{tab:stmVSpulpV3_energy} shows the performance and power measurements of the HD computing on the PULPv3 prototype~\cite{PULPv3} in different operating conditions, and compares it with the ARM Cortex M4, benchmarked on an STM32F4-DISCOVERY board.
In this experiment we use 10,000-D to retain to the best accuracy of 92.4\%, and accordingly configure the clock frequency of the processors to achieve a detection latency of 10\,ms~\cite{benatti2017prosthetic,1710164,khezri2007real}.  
The second column of Table~\ref{tab:stmVSpulpV3_energy} shows that with respect to the single-core PULPv3, the ARM Cortex M4 can operate at a lower frequency for the target detection latency, exploiting some optimized instructions that speed up the execution, namely \textit{load and shift} and \textit{load 32-bit immediate}.
The key features of the PULPv3 SoC exploited in this work are performance-tunable near-threshold computing and parallelism.
The 4.9$\times$ power gap between the ARM Cortex M4 and the single-core PULPv3 power at 0.7\,V is partially given by the technology gap (i.e., 90\,nm vs. 28\,nm), but also by the cluster architecture and its implementation strategy optimized for energy-efficient operation~\cite{PULPv3}.
Significant energy boost can be achieved through parallel computing over the 4 cores of the cluster.
This allows to fully exploit the parallel compute power of the cluster and to reduce the operating frequency of the system by 3.72$\times$ (almost ideal speed-up over 4 cores), which saves significant power, leading to 8.1$\times$ power reduction with respect to the ARM Cortex M4, at the operating voltage of 0.7V.
Finally we exploit the process and temperature compensation capabilities of the SoC to enable aggressive voltage scaling, still reaching the target operating frequency of 14.3\,MHz~\cite{PULPv3}. 
This key feature of the PULPv3 SoC allows to scale the voltage of the cluster down to 0.5\,V, improving energy efficiency and leading to a power reduction of about one order of magnitude (9.9$\times$) with respect to the ARM Cortex M4.

It should be noted that the clock generation subsystem of PULPv3, composed of 2 frequency locked loops (FLL), is not optimized for low-power operation, featuring a reference frequency of 40\,MHz and a power consumption of 1,45\,mW. 
This block forms a bottleneck for energy efficiency at low voltage, dominating the overall power of the system. Replacing this block with a new generation FLL optimized for low-power~\cite{BellasiFLL} would reduce the clock generation power by 4$\times$ leading to a further 2$\times$ reduction of system power, and boosting energy efficiency by $\approx$20$\times$ with respect to the ARM Cortex M4. 
This result motivates us to assess the accelerator with larger workloads, and devise further architectural optimizations.
\section{Improved Accelerator and Scalability}
\label{section:discussion}
This section describes how the performance of HD computing can be optimized on a new generation PULP platform (Wolf) featuring an optimized cluster architecture~\cite{fulmine} and RISC-V processors enhanced with ISA extensions targeting energy efficient digital signal processing~\cite{gautschi2017near}.
We also show how our accelerator allows to increase the workload of the processing chain without exceeding a 10\,ms detection latency, which is an order of magnitude lower than state-of-the-art systems~\cite{benatti2017prosthetic,1710164,khezri2007real}.
\subsection{HD Computing on Wolf}
With respect to the PULPv3 architecture, the main architectural improvement of the Wolf cluster include a better scalability (up to 8 processors), an hardware synchronization mechanism which allows to significantly reduce the programming overheads of the OpenMP runtime, fully exploiting the intrinsic parallelism of applications, and an enhanced processor extending the RISC-V ISA with advanced arithmetic operations, that can be inserted in optimized C code adopting built-in functions~\cite{gautschi2017near}. 
The flavor of the dedicated instructions that can be exploited in HD computing mainly include those accelerating \textit{for} loops and bit manipulation instructions. 
Indeed, in the processing chain of HD, there are several operations where single bits need to be read/inserted from/into 32-bit words, and where the number of 1's in a 32-bit word needs to be counted.
\begin{table}[t]
  \caption{Performance of accelerated HD computing on PULPv3 versus Wolf. Results refer to the execution with built-in, 10,000-D, N=1; Cyc, ld, sp stand for cycles, load, and speed-up (sp wrt PULPv3 1 core).}
  \label{tab:table_perf_builtin_n1}
  \scalebox{0.58}{
  \begin{tabular}{lrrrcrcrcrrc}
    \toprule
    & \multicolumn{2}{c}{\textbf{PULPv3}}  & \multicolumn{2}{c}{\textbf{PULPv3}} & \multicolumn{2}{c}{\textbf{Wolf}} & \multicolumn{2}{c}{\textbf{Wolf}}  & \multicolumn{3}{c}{\textbf{Wolf}} \\
 & \multicolumn{2}{c}{\textbf{1 core}}  & \multicolumn{2}{c}{\textbf{4 cores}} & \multicolumn{2}{c}{\textbf{1 core}} & \multicolumn{2}{c}{\textbf{1 core built-in}}  & \multicolumn{3}{c}{\textbf{8 cores built-in}} \\
    \midrule
    \textbf{Kernel}  & cyc(k) &ld(\%) & cyc(k)  & \textbf{sp($\times$)} & cyc(k) & \textbf{sp($\times$)} & cyc(k) & \textbf{sp($\times$)} & cyc(k) &ld(\%) & \textbf{sp($\times$)} \\
    \texttt{MAP+ENCODERS}   &492 &92.30     &129   &\textbf{3.81} &401  &\textbf{1.23}  &176  &\textbf{2.80}  &25 &86.21  &\textbf{19.68}  \\
    \texttt{AM}  	        &41  &7.70      &14    &\textbf{2.93} &33   &\textbf{1.24}  &12   &\textbf{3.42}  &4  &13.79  &\textbf{10.25}  \\
    \texttt{TOTAL}          &533 &100.00    &143   &\textbf{3.73} &434  &\textbf{1.23}  &188  &\textbf{2.84}  &29 &100.00 &\textbf{18.38}  \\
    \bottomrule
  \end{tabular}
  }
  \vspace{-5mm}
\end{table}

The bit manipulation instructions used to optimize the performance of the application are \textit{p.extractu}, \textit{p.insert} and \textit{p.cnt}. 
The first and the second built-ins, \textit{p.extractu} and \textit{p.insert}, are used respectively to read and set the value assumed by a given bit in an unsigned 32-bit integer variable in a register.
The last one, \textit{p.cnt}, is so-called \textit{popcount} and gives the number of bits set to 1 in a word. 
The built-ins \textit{p.extractu} and \textit{p.insert} are used to further optimize the spacial encoder kernel. 
In this part of the processing chain after binding each channel to its signal level in the HD space, a componentwise majority is needed to be applied on these bound hypervectors to produce the spatial hypervector, i.e., $S^t=[(E_1 \oplus V_1^t) + ... + (E_i \oplus V_i^t)]$. 
As shown in Fig. \ref{fig:popenc} (left), the componentwise majority operation needs to extract $i$ components of these hypervectors (i.e., bit-by-bit) and to count the number of bits that are set to 1 for the majority voting. 
If the number of channels ($i$) is even, one random but \emph{reproducible} hypervector is generated, by componentwise XOR between two bound hypervectors, for the majority to break the ties at random.
For instance, with four channels, we use five bound hypervectors for the majority, and extract and insert five bits (one bit from every hypervector) in an unsigned integer. 
Then, we use the \textit{popcount} (\textit{p.cnt}) to decide whether the number of bits that are set to 1 is higher than the number of bits that are set to 0. 
If it holds, we set the related bit (i.e., the same component) to 1 in the spatial hypervector.   

The \textit{popcount} is used in the AM kernel as well. 
Here, the Hamming distances between the query hypervector and the prototype hypervectors stored in the AM matrix are computed. 
To do that, the \textit{popcount} is applied to all variables that compose the hypervectors after the componentwise XOR operation between the query and the prototype hypervectors.
Table~\ref{tab:table_perf_builtin_n1} shows that 1.23$\times$ speed-up is achieved by migrating from the single-core PULPv3 to the single-core Wolf architecture with a general-purpose ANSI-C code, thanks to the optimized RISC-V ISA and compiler. Further 2.3x speedup can be achieved on the Wolf SoC thanks to the support of the specialized instruction extensions that are included in the C code as built-ins (2.8$\times$ wrt single-core PULPv3).

Moreover, our accelerator linearly benefits from larger number of cores that are available in Wolf.
Table~\ref{tab:table_perf_builtin_n1} also summarizes the execution time in clock cycles of the HD Computing using various number of cores in PULPv3 and Wolf. 
When a larger number of cores is used, a conspicuous reduction in the execution time is achieved. 
These results show that the accelerator can scale perfectly among multiple cores (up to 8 cores). In fact, a speed-up of 3.7$\times$ is obtained by moving execution from single-core to 4 cores on PULPv3, while the implementation on the Wolf cluster gains 6.5$\times$ speedup, scaling from single-core to 8 cores.   
The map and encoding kernels present nearly ideal speed-ups, while the AM kernel tends to saturate the improvement.
The main reason is that the computational load is small and the OpenMP runtime overhead increasingly degrades the parallel performance. 
Despite this, the impact on the total gain is negligible. 
As shown in Table~\ref{tab:table_perf_builtin_n1}, the single-core and 8-core Wolf are around 2.8$\times$ and 18.4$\times$ faster compared to the single-core PULPv3.  
These excellent improvements in our accelerator is achieved as a cumulative results of the better ISA, compiler, as well as the built-in extensions on the 8-core Wolf cluster.
In the single-core PULPv3, the map and encoding kernels require 92.30\% of the overall execution, while the AM kernel takes the remaining computational load (7.70\%). In the 8-core Wolf execution with built-in this gap decreases as a result of the saturation in speed-up due to the OpenMP runtime overhead.
Hereafter, the exploration and scalability analysis are done on the Wolf. 
\begin{figure}[t]
\begin{center}
    \includegraphics[width=0.89\columnwidth]{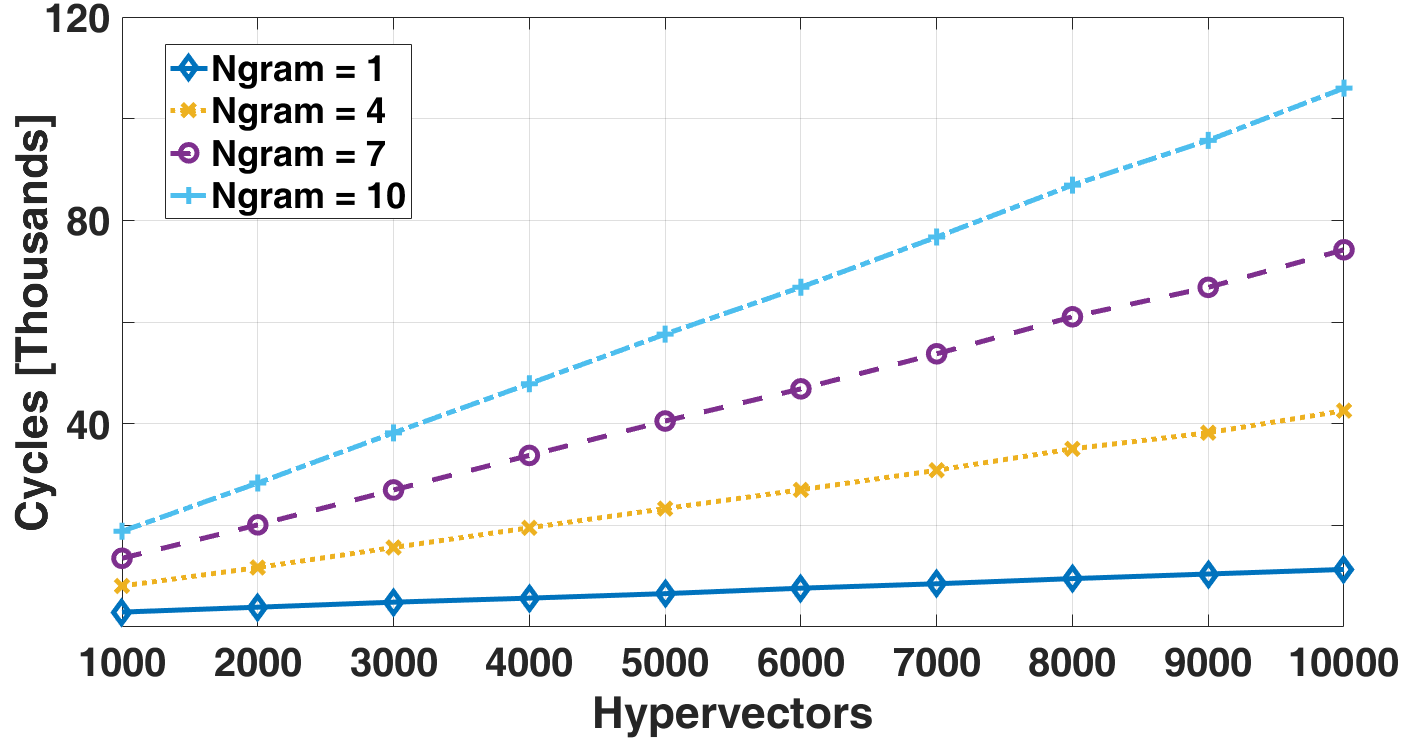}
\caption{Exploring the dimension of the hypervectors with different large N-grams on Wolf 8 cores with built-in.}
\label{fig:graph_hv_vs_cycles}
\vspace{-5mm}
\end{center}
\end{figure}
\subsection{Accelerator Scalability}
\begin{figure}[t]
\begin{center}
    \includegraphics[width=0.86\columnwidth]{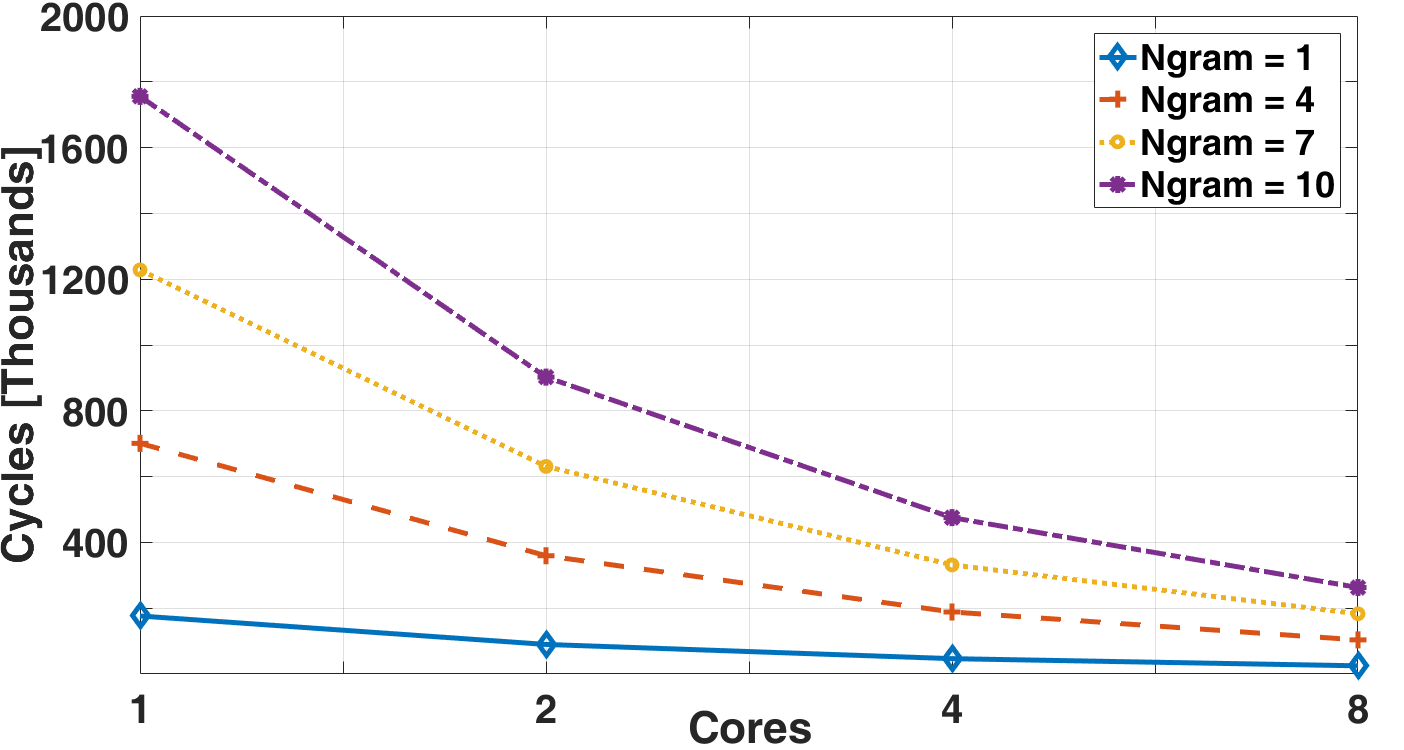}
\caption{Performance of accelerated HD computing with large N-grams when executing on multiple cores on Wolf with built-in and 10,000-D.}
\label{fig:graph_core_vs_cycles}
\vspace{-4mm}
\end{center}
\end{figure}

\begin{figure}[t]
\begin{center}
   \includegraphics[width=0.87\columnwidth]{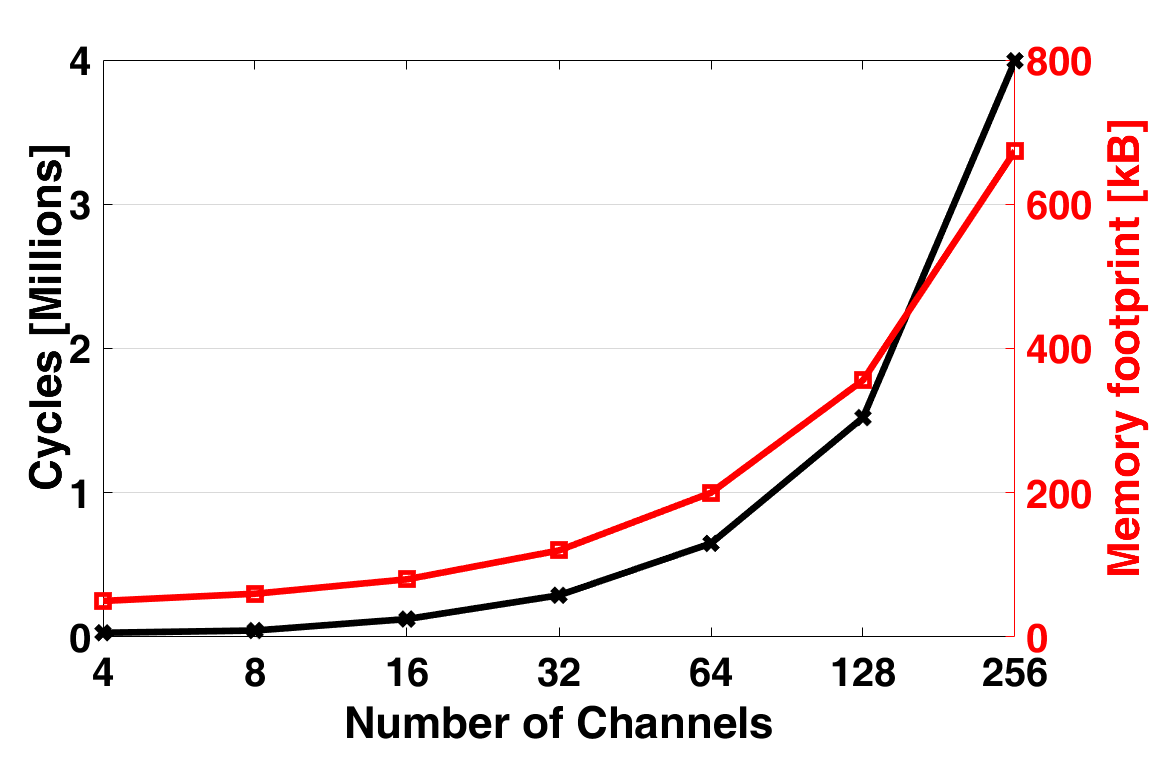}\label{fig:rmsval} 
  \caption{Performance and memory footprint of accelerated HD computing with increased number of channels on 8 cores Wolf with built-in and 10,000-D .}
  \label{fig:scale_chs_memory}
\end{center}
\vspace{-4mm}
\end{figure}

The earlier results presented in Section~\ref{section:system_description} focused on the EMG task with a small number of four channels and an N-gram size of one.
However, as we mentioned, for more complex tasks such as EEG classification, a larger number of channels and wider temporal window (i.e., larger N-gram size) are required~\cite{BICT17}.
Therefore, we assess the scalability of our accelerator by extensively increasing the number of channels up to 256,  the N-gram size up to 10 and the dimension of hypervectors up to 10000, showing that the accelerator can be tailored for other type of applications.
The dimensionality is related to capacity of hypervectors. 
Increasing the dimensionality of the hypervectors creates higher capacity for handling more complex tasks, and leads to an overall increase in the number of operations in the processing chain. 
Fig.~\ref{fig:graph_hv_vs_cycles} demonstrates that increasing the dimension of the hypervectors, for every N-gram size, corresponds to a linear growth of the execution time in terms of number of clock cycles. 
Hereinafter, the dimension of the hypervectors are fixed to 10,000-D to explore the capability of the accelerator with higher computational requirements for complex tasks.
We also evaluate how increasing the size of N-gram from 1 to 10 affects the performance of our accelerator, and how this workload scales among different cores. 
Fig.~\ref{fig:graph_core_vs_cycles} demonstrates that the accelerator is able to scale such excessive workload perfectly among the cores.   
As shown in Fig.~\ref{fig:scale_chs_memory}, ranging the number of channels  from 4 to 256,  the clock cycles increases linearly with the number of channels, and our accelerator meets the latency constraint. 
We should note that the commercial ARM Cortex M4 could not handle such a workload for HD computing: it cannot meet the 10\,ms latency constraint when the number of channels is larger than 16. 
Moreover, a linear increase in the number of channels induces only a linear growth of the the memory footprint (Fig.~\ref{fig:scale_chs_memory}, red line) to store and allocate all the matrices for the HD computing. 
This is a superb property of HD computing as the memory footprint has a considerable impact on the design of the embedded ultra-low power architectures.    

\section{Conclusion}
This work presents accelerating HD computing on the PULP platform with optimized operations and memory accesses.
We show its application on the EMG hand gesture classification that surpasses the state-of-the-art SVM accuracy.
For the end-to-end execution of classification, our accelerator in PULPv3 achieves 3.7$\times$ speed-up and 2$\times$ energy saving compared to its single-core execution; it also achieves 9.9$\times$ energy saving compared to the ARM Cortex M4.
We further evaluate our accelerator in Wolf that demonstrates nearly ideal speed-up by exploiting bit-manipulation ISA extensions and larger cores: Wolf with single core and 8 cores achieves 2.8$\times$ and 18.4$\times$ faster execution compared to the single-core PULPv3.
Moreover, we show that increasing the number of input channels to 256, and the length of temporal window to 10 (i.e., the N-gram size) poses only a linear growth in the execution time and the memory footprint that are efficiently handled by our accelerator without exceeding the 10\,ms detection latency requirement. 

\section*{Acknowledgment}
This work was supported by the European project EuroCPS (grant n. 644090), the ETH Zurich Postdoctoral Fellowship and the Marie Curie Actions for People COFUND programs.

\bibliographystyle{ACM-Reference-Format}

\bibliography{bibliography.bbl} 

\end{document}